\documentclass{aastex61}

\received{}
\revised{}
\accepted{}
\submitjournal{ApJ}

\shorttitle{Survey Observations of HC$_{3}$N and HC$_{5}$N}
\shortauthors{Taniguchi et al.}

\begin{document}

\title{Survey Observations to Study Chemical Evolution from High-Mass Starless Cores to High-Mass Protostellar Objects I: HC$_{3}$N and HC$_{5}$N}

\correspondingauthor{Kotomi Taniguchi}
\email{kotomi.taniguchi@nao.ac.jp}

\author{Kotomi Taniguchi}
\altaffiliation{Research Fellow of Japan Society for the Promotion of Science}
\affiliation{Department of Astronomical Science, School of Physical Science, SOKENDAI (The Graduate University for Advanced Studies), Osawa, Mitaka, Tokyo 181-8588, Japan}
\affiliation{Nobeyama Radio Observatory, National Astronomical Observatory of Japan, Minamimaki, Minamisaku, Nagano 384-1305, Japan}

\author{Masao Saito}
\altaffiliation{The present address : National Astronomical Observatory of Japan, Osawa, Mitaka, Tokyo 181-8588, Japan}
\affiliation{Nobeyama Radio Observatory, National Astronomical Observatory of Japan, Minamimaki, Minamisaku, Nagano 384-1305, Japan}
\affiliation{Department of Astronomical Science, School of Physical Science, SOKENDAI (The Graduate University for Advanced Studies), Osawa, Mitaka, Tokyo 181-8588, Japan}

\author{T. K. Sridharan}
\affiliation{Harvard-Smithsonian Center for Astrophysics, 60 Garden Street, MS 78, Cambridge, MA 02138, USA}

\author{Tetsuhiro Minamidani}
\affiliation{Nobeyama Radio Observatory, National Astronomical Observatory of Japan, Minamimaki, Minamisaku, Nagano 384-1305, Japan}
\affiliation{Department of Astronomical Science, School of Physical Science, SOKENDAI (The Graduate University for Advanced Studies), Osawa, Mitaka, Tokyo 181-8588, Japan}



\begin{abstract}

We carried out survey observations of HC$_{3}$N and HC$_{5}$N in the 42$-$45 GHz band toward 17 high-mass starless cores (HMSCs) and 35 high-mass protostellar objects (HMPOs) with the Nobeyama 45 m radio telescope.
We have detected HC$_{3}$N from 15 HMSCs and 28 HMPOs, and HC$_{5}$N from 5 HMSCs and 14 HMPOs, respectively.
The average values of the column density of HC$_{3}$N are found to be ($5.7 \pm 0.7$)$\times 10^{12}$ and ($1.03 \pm 0.12$)$\times 10^{13}$ cm$^{-2}$ in HMSCs and HMPOs, respectively. 
The average values of the fractional abundance of HC$_{3}$N are derived to be ($6.6 \pm 0.8$)$\times 10^{-11}$ and ($3.6 \pm 0.5$)$\times 10^{-11}$ in HMSCs and HMPOs, respectively.  
We find that the fractional abundance of HC$_{3}$N decreases from HMSCs to HMPOs using the Kolmogorov-Smirnov test. 
On the other hand, its average value of the column density slightly increases from HMSCs to HMPOs.
This may imply that HC$_{3}$N is newly formed in dense gas in HMPO regions.
We also investigate the relationship between the column density of HC$_{3}$N in HMPOs and the luminosity-to-mass ratio ($L/M$), a physical evolutional indicator.
The column density of HC$_{3}$N tends to decrease with increasing the $L/M$ ratio, which suggests that HC$_{3}$N is destroyed by the stellar activities.

\end{abstract}




\section{Introduction} \label{sec:intro}

Our understanding of the high-mass star-forming regions, where massive stars ($M \geq 8$ M$_{\sun}$) are born, is still poor, in contrast to the progress of studies in the low-mass star-forming regions.
In particular, the chemical evolution in the high-mass star-forming regions is not still clear yet.
There are only a handful of such studies.
Recently, MALT90 survey \citep{2011ApJS..197...25F, 2013PASA...30...57J} using the Australia Telescope National Facility Mopra 22 m single-dish telescope provided us with the large sample data of molecules in the 90 GHz band. 
\citet{2013ApJ...777..157H} investigated the chemical evolution in the high-mass star-forming regions, and found that N$_{2}$H$^{+}$ and HCO$^{+}$ abundances increase as a function of evolutionary stage.
They also found that the $I$[HCN (1$-$0)]/$I$[HNC (1$-$0)] integrated intensity ratios marginally showed evidence of an increase as the clumps evolve.
\citet{2015MNRAS...451...2507Y} investigated massive extended green objects (EGOs), and suggested that N$_{2}$H$^{+}$ and C$_{2}$H might be used as "chemical clocks" by comparing with other molecules such as H$^{13}$CO$^{+}$.
\citet{2014A&A...563A..97G} carried out observations toward 59 high-mass star-forming regions, which are categorized into infrared dark clouds (IRDCs), high-mass protostellar objects (HMPOs), hot molecular cores (HMCs), and ultracompact \ion{H}{2} regions (UC\ion{H}{2}), using the IRAM 30 m telescope.
They found possibilities that the chemical composition varies along with the evolutionary stages.

In the low-mass star-forming regions, carbon-chain molecules are well known as good chemical evolutional tracers \citep{1992ApJ...392..551S, 2009ApJ...699..585H}.
Carbon-chain molecules account for approximately 40\% of the $\sim 200$ molecules detected in the interstellar medium and circumstellar shells.
They are efficiently formed by the gas-phase ion-molecule reactions involving carbon atoms (C) or carbon cations (C$^{+}$) in young  dark clouds.
Carbon-chain species decrease by depletion onto dust grains, destruction by the UV radiation, and reactions with oxygen atoms \citep{2013ChRv...113...8981}.
Hence, carbon-chain molecules are abundant in young low-mass starless cores, while they are deficient in evolved low-mass star-forming cores.
Moreover, in some low-mass starless cores, the main formation mechanisms of some carbon-chain molecules were investigated using their $^{13}$C isotopic fractionation \citep[e.g.,][]{2016ApJ...817..147T, 2017ApJ...846...46T} and $^{15}$N isotopic fractionation \citep{2017PASJ...69L...7T}.
On the other hand, there are few studies about the carbon-chain chemistry in the high-mass star-forming regions.

Several studies similar to ones towards the low-mass star-forming regions investigating the chemical evolution of carbon-chain molecules have been conducted toward the giant molecular clouds (GMCs) (Orion A GMC; \citet{2010PASJ...62.1473T, 2014PASJ...66...16T, 2014PASJ...66..119O} and Vela C GMC; \citet{2016PASJ...68....3O}).
However, their core samples contained the low-mass cores with a few M$_ {\sun}$ $-$ a dozen M$_ {\sun}$, which are not massive enough to form massive stars.
Therefore, their studies were not enough to confirm the chemical evolution of carbon-chain molecules in the high-mass star-forming regions.

\citet{2008ApJ...678...1049S} carried out observations toward 55 IRDCs with the Nobeyama 45 m radio telescope and the Atacama Submillimeter Telescope Experiment 10 m telescope.
They detected HC$_{3}$N and NH$_{3}$ toward almost of the sources, while CCS was not detected toward any of the sources.
\citet{2008ApJ...678...1049S} suggested that most of the massive clumps were chemically evolved than the low-mass starless cores from their survey results.
However, they did not investigate in detail the relationships between the column density of carbon-chain molecules and N-bearing species (NH$_{3}$ and N$_{2}$H$^{+}$), which are the late-type species, as shown in \citet{1992ApJ...392..551S, 2009ApJ...699..585H}.
Therefore, there is no clear result about the chemical evolution of carbon-chain molecules in the high-mass star-forming regions.

In general, carbon-chain molecules are thought to be deficient in hot core regions \citep[e.g.,][]{2009ARA&A..47..427H}.
However, there were several chemical network model simulations showing that cyanopolyyne series (HC$_{2n+1}$N) could be formed and survive in hot core regions \citep{2009MNRAS.394..221C, 2011ApJ...743..182H}.
Several observations of cyanopolyynes toward hot cores have been carried out.
From the $^{13}$C isotopic fractionation of HC$_{3}$N, \citet{2016ApJ...830..106T} proposed that the main formation pathway of HC$_{3}$N is the reaction between C$_{2}$H$_{2}$ and CN in the G28.28-0.36 hot core.
Their suggestion is consistent with the chemical network simulation conducted by \citet{ 2009MNRAS.394..221C}.
\citet{2014MNRAS.443.2252G} carried out a survey observation of HC$_{5}$N, which was though to be deficient in hot cores, toward 79 hot cores associated with the 6.7 GHz methanol masers, and detected it from 35 hot cores.
In addition, \citet{2017ApJ...844...68T} carried out observations toward four hot cores where \citet{2014MNRAS.443.2252G} detected HC$_{5}$N.
They found that HC$_{5}$N exists in the warm gas around the massive young stellar objects and derived its considerable fractional abundances.
These results suggest that HC$_{5}$N can exist even at the hot core stage and mean that we can investigate the chemical evolution of carbon-chain molecules in the high-mass star-forming regions.

In the present paper, we report the survey observations of HC$_{3}$N and HC$_{5}$N in the 42$-$45 GHz band toward 17 high-mass starless cores (HMSCs) and 35 high-mass protostellar objects (HMPOs) using the Nobeyama 45 m radio telescope.
Our target sources were selected from the HMSC list \citep{2005ApJ...634L..57S} and the HMPO list \citep{2002ApJ...566..931S}.
\citet{2002ApJ...566..931S} identified 69 HMPOs, using far-infrared, radio continuum, and molecular data.
Millimeter dust continuum emission was detected from all of the sources \citep{2002ApJ...566..945B}, whereas most of the sources showed weak or no continuum emission at 3.6 cm.
\citet{2005ApJ...634L..57S} identified 56 HMSCs, comparing images of fields containing candidate HMPOs at 1.2 mm and mid-infrared (MIR; 8.3 \micron).
HMSC was defined as a core showing 1.2 mm emission and absorption or no emission at the MIR wavelength suggestive of cold dust.
HMSCs are thought to be in an early stage with no signs of massive star formation, before the HMPO phase \citep{2005ApJ...634L..57S}.

\section{Observations} \label{sec:obs}

We carried out observations of HC$_{3}$N ($J=5-4$; 45.49031 GHz, \citet{2005JMoSt...742...215M}) and HC$_{5}$N ($J=16-15$; 42.60215 GHz, \citet{2005JMoSt...742...215M}) simultaneously with the Nobeyama 45 m radio telescope in 2015 December, 2016 May and June (2015-2016 season).
Our target sources were 17 HMSCs and 35 HMPOs, which were selected from the HMSC list \citep{2005ApJ...634L..57S} and the HMPO list \citep{2002ApJ...566..931S}, as listed in Tables \ref{tab:tab1} and \ref{tab:tab2}.
The selected sources have the following characteristics:
\begin{enumerate}
\item the source declination is above -6\arcdeg for HMSCs and +6\arcdeg for HMPOs,
\item NH$_{3}$ has been detected, and
\item HMPOs located in the same regions as the observed HMSCs (-6\arcdeg $<$ declination $<$ +6\arcdeg).
\end{enumerate}
The exact observed positions were listed in \citet{2002ApJ...566..931S, 2005ApJ...634L..57S}.

We used the Z45 receiver \citep{2015PASJ...67..117N}, which enables us to obtain dual-linear-polarization data simultaneously.
The main beam efficiency ($\eta_{\rm {B}}$) and the beam size (HPBW) were 72\% and 37$^{\prime\prime}$, respectively.
The system temperatures were from 110 to 220 K, depending on the weather conditions and elevations.
We used the SAM45 FX-type digital correlator \citep{2012PASJ...64...29K} in frequency setting whose bandwidth and resolution are 250 MHz and 61.04 kHz, respectively.
The frequency resolution corresponds to 0.4 km s$^{-1}$.
The standard chopper-wheel calibration method was employed to convert the receiver output intensity into {\it T}$^{\ast}_{\rm A}$ and uncertainty in the absolute scale is expected to be approximately 10\%.

The telescope pointing was checked every 1.5$-$3 hr depending on the wind conditions with the Z45 receiver by observing the SiO maser lines ($J=1-0$) from U-Aur, RR-Aql, R-Aql, UX-Cyg, IRC+60334, and R-Cas.
The pointing error was less than 3$^{\prime\prime}$.
We employed the position-switching mode.
We set the off-source positions at no IRAS 100 $\micron$ emission positions\footnote{We used the SkyView (http://skyview.gsfc.nasa.gov/current/cgi/query.pl).} or low extinction ($A_{V} < 1$ mag) positions\footnote{We used the all-sky visual extinction map generated by \citet{2005PASJ...57S...1D} (http://darkclouds.u-gakugei.ac.jp).}. 
The smoothed bandpass calibration method \citep{2012PASJ...64..118Y} was adopted in data reduction.
We set on-source and off-source integration time at 20 and 5 seconds, respectively.
We applied 16-channel smoothing for off-source data to reduce their noise level.
We conducted 2-channel binning in the final spectra.

\section{Results and Analysis}

\subsection{Results} \label{sec:res}

We conducted data reduction using the Java NEWSTAR\footnote{http://www.nro.nao.ac.jp/$\sim$jnewstar/html/}, which is software for data reduction and analyses of the Nobeyama data.
Figures \ref{tab:tab1} $-$ \ref{tab:tab3} show the spectra of HC$_{3}$N (upper) and HC$_{5}$N (lower) in all of the HMSCs and HMPOs.
We fitted the spectra with a Gaussian profile, and obtained spectral line parameters, as summarized in Tables \ref{tab:tab1} and \ref{tab:tab2}.
The absolute peak values may include another 10\% calibration uncertainty.
HC$_{3}$N was detected from 15 HMSCs and 28 HMPOs with the signal-to-noise ratios above 4 in {\it T}$^{\ast}_{\rm A}$ scale.
We checked the 1.2 mm continuum images \citep{2002ApJ...566..945B}, and found that HC$_{3}$N was not detected only when the beam did not cover the 1.2 mm continuum cores.
The detection does not depend on the source distances.
In HMSC 18454-0158-9, we applied two-component Gaussian fitting.
The spectra of HC$_{3}$N in some HMPOs show wing emissions suggestive of the molecular outflow shock origin.
These results are consistent with the suggestion by \citet{2015ApJS..221...31S} that HC$_{3}$N is an outflow shock origin species in OMC-2.

\begin{figure}
\figurenum{4}
\plotone{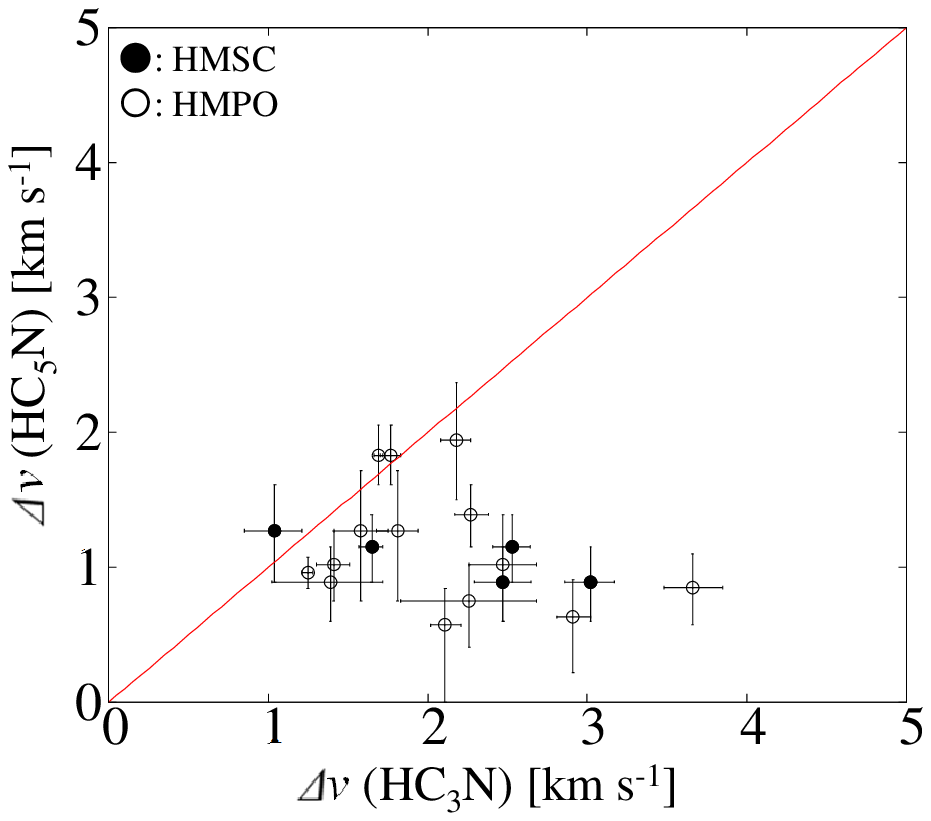}
\caption{Plot of the FWHM between HC$_{3}$N vs. HC$_{5}$N. The errors were estimated from the Gaussian fitting errors and the velocity resolution. The red line shows $\Delta v$(HC$_{3}$N) $=$ $\Delta v$(HC$_{5}$N). \label{fig:f9}}
\end{figure}

HC$_{5}$N was detected from  5 HMSCs and 14 HMPOs with the signal-to-noise ratios above 4 in {\it T}$^{\ast}_{\rm A}$ scale.
Figure \ref{fig:f9} shows the relationship of the line width between HC$_{3}$N and HC$_{5}$N.
The line widths ($\Delta v$) are calculated using the following formula:

\begin{equation} \label{FWHM}
\Delta v = \sqrt{\Delta v_{\rm {obs}}^2 - \Delta v_{\rm {inst}}^2},
\end{equation}
where $\Delta v_{\rm {obs}}$ and $\Delta v_{\rm {inst}}$ are the observed line width (Tables \ref{tab:tab1} and \ref{tab:tab2}) and the instrumental velocity width (0.8 km s$^{-1}$, Section \ref{sec:obs}), respectively.
Although the HC$_{5}$N line width estimates are less certain as they are comparable to the spectral resolution, it is clear that the HC$_{3}$N lines are significantly broader. 
These results suggest that HC$_{5}$N exists only in quiescent gas with less internal motions, while HC$_{3}$N exists in more active regions.
The values of $V_{\rm {LSR}}$ of HC$_{3}$N and HC$_{5}$N are consistent with those summarized in \citet{2002ApJ...566..931S,2005ApJ...634L..57S}, except for HMSC 18385-0512-3.
 
\floattable
\begin{deluxetable}{lccccccccccc}
\tabletypesize{\scriptsize}
\tablecaption{Spectral Line Parameters in HMSCs\label{tab:tab1}}
\tablewidth{0pt}
\tablehead{
\colhead{} & \multicolumn{5}{c}{HC$_{3}$N ($J=5-4$)} & \colhead{} & \multicolumn{5}{c}{HC$_{5}$N ($J=16-15$)} \\
\cline{2-6}\cline{8-12}
\colhead{Source} & \colhead{{\it T}$^{\ast}_{\mathrm A}$} & \colhead{$\Delta v$\tablenotemark{a}} & \colhead{{\it V}$_{\mathrm {LSR}}$\tablenotemark{b}} & \colhead{$\int T^{\ast}_{\mathrm A}dv$} & \colhead{rms\tablenotemark{c}} & \colhead{} & \colhead{{\it T}$^{\ast}_{\mathrm A}$} & \colhead{$\Delta v$\tablenotemark{a}} & \colhead{{\it V}$_{\mathrm {LSR}}$\tablenotemark{b}} & \colhead{$\int T^{\ast}_{\mathrm A}dv$} & \colhead{rms\tablenotemark{c}} \\
\colhead{} & \colhead{(K)} & \colhead{(km s$^{-1}$)} & \colhead{(km s$^{-1}$)} & \colhead{(K km s$^{-1}$)} & \colhead{(mK)} &\colhead{} & \colhead{(K)} & \colhead{(km s$^{-1}$)} & \colhead{(km s$^{-1}$)} & \colhead{(K km s$^{-1}$)} & \colhead{(mK)}
}
\startdata
18385-0512-3 & 0.398 (14) & 1.83 (7) & 46.6 & 0.78 (4) & 8.8 & & 0.046 (9) & 1.4 (2) & 46.9 & 0.07 (2) & 8.6 \\
18437-0216-3 & 0.088 (12) & 2.2 (3) & 110.2 & 0.21 (4) & 13.2 & & $< 0.03$ ($3\sigma$) & ...& ... & ... & 9.5 \\
18445-0222-4 & 0.045 (10) & 1.3 (3) & 88.6 & 0.06 (2) & 9.8 & & $< 0.03$ ($3\sigma$) & ... & ... & ... & 9.9 \\
18447-0229-3 & $< 0.03$ ($3\sigma$) & ... & ... & ... & 8.9 & & $< 0.03$ ($3\sigma$) & ... & ... & ... & 8.9 \\
18447-0229-4 & 0.112 (6) & 2.60 (17) & 99.5 & 0.31 (3) & 7.1 & & 0.047 (7) & 1.2 (2) & 99.4 & 0.062 (14) & 7.1 \\
18447-0229-5 & 0.094 (9) & 1.32 (13) & 104.7 & 0.13 (2) & 8.6 & & $< 0.03$ ($3\sigma$) & ... & ... & ... & 9.4 \\
18454-0158-1 & 0.157 (10) & 1.64 (12) & 100.5 & 0.27 (3) & 10.2 & & $< 0.03$ ($3\sigma$) & ... & ... & ... & 9.9 \\
18454-0158-3 & 0.453 (9) & 4.17 (17) & 97.9 & 2.01 (9) & 8.7 & & $< 0.03$ ($3\sigma$) & ... & ... & ... & 8.9 \\
18454-0158-5 & 0.411 (9) & 3.12 (15) & 94.1 & 1.36 (7) & 8.7 & & 0.054 (9) & 1.2 (2) & 95.4 & 0.067 (16) & 8.7 \\
18454-0158-8 & 0.147 (11) & 2.7 (2) & 95.6 & 0.42 (5) & 12.4 & & $< 0.03$ ($3\sigma$) & ... & ...& ... & 10.2 \\
18454-0158-9 & 0.426 (10) & 4.2 (2) & 96.3 & 1.90 (10)& 9.6 & & $< 0.03$ ($3\sigma$) & ...& ... & ... & 7.9 \\
                       & 0.239 (10) & 4.8 (2) & 100.9 & 1.22 (7) & 9.6 & & $< 0.03$ ($3\sigma$) & ... & ... & ... & 7.9  \\
18454-0158-10 & $< 0.03$ ($3\sigma$) & ... & ... & ... & 9.6 & & $< 0.02$ ($3\sigma$) & ... & ... & ... & 7.7 \\
19175+1357-3 & 0.080 (10) & 2.0 (2) & 7.3 & 0.17 (3) & 9.7 & & $< 0.03$ ($3\sigma$) & ... & ... & ... & 8.9 \\
19175+1357-4 & 0.117 (8) & 1.60 (13) & 7.3 & 0.20 (2) & 8.7 & & $< 0.03$ ($3\sigma$) & ... & ... & ... & 8.6 \\
19410+2336-2 & 0.279 (10) & 2.65 (11) & 21.3 & 0.79 (4) & 11.2 & & 0.062 (8) & 1.4 (2) & 21.4 & 0.092 (18) & 8.3 \\
20081+2720-1 & 0.096 (10) & 1.31 (14) & 5.0 & 0.13 (2) & 9.7 & & 0.036 (8) & 1.5 (3) & 4.1 & 0.059 (19) & 7.9 \\
22570+5912-3 & 0.035 (5) & 4.2 (7) & -46.1 & 0.15 (3) & 8.3 & & $< 0.03$ ($3\sigma$) & ... & ... & ... & 9.2 \\
\enddata
\tablecomments{The numbers in parentheses represent one standard deviation in the Gaussian fit. The errors are written in units of the last significant digit.}
\tablenotetext{a}{These values are not corrected for instrumental velocity resolution.}
\tablenotetext{b}{The error of $V_{\rm {LSR}}$ is commonly 0.8 km s$^{-1}$, corresponding to the velocity resolution of the final spectra (Section \ref{sec:obs}).}
\tablenotetext{c}{The rms noises in emission-free region.}
\end{deluxetable}

\floattable
\begin{deluxetable}{lccccccccccc}
\tabletypesize{\scriptsize}
\tablecaption{Spectral Line Parameters in HMPOs\label{tab:tab2}}
\tablewidth{0pt}
\tablehead{
\colhead{} & \multicolumn{5}{c}{HC$_{3}$N ($J=5-4$)} & \colhead{} & \multicolumn{5}{c}{HC$_{5}$N ($J=16-15$)} \\
\cline{2-6}\cline{8-12}
\colhead{Source} & \colhead{{\it T}$^{\ast}_{\mathrm A}$} & \colhead{$\Delta v$\tablenotemark{a}} & \colhead{{\it V}$_{\mathrm {LSR}}$\tablenotemark{b}} & \colhead{$\int T^{\ast}_{\mathrm A}dv$} & \colhead{rms\tablenotemark{c}} & \colhead{} & \colhead{{\it T}$^{\ast}_{\mathrm A}$} & \colhead{$\Delta v$\tablenotemark{a}} & \colhead{{\it V}$_{\mathrm {LSR}}$\tablenotemark{b}} & \colhead{$\int T^{\ast}_{\mathrm A}dv$} & \colhead{rms\tablenotemark{c}} \\
\colhead{} & \colhead{(K)} & \colhead{(km s$^{-1}$)} & \colhead{(km s$^{-1}$)} & \colhead{(K km s$^{-1}$)} & \colhead{(mK)} &\colhead{} & \colhead{(K)} & \colhead{(km s$^{-1}$)} & \colhead{(km s$^{-1}$)} & \colhead{(K km s$^{-1}$)} & \colhead{(mK)}
}
\startdata
05358+3543 & 0.477 (16) & 2.26 (9) & -17.9 & 1.15 (6) & 9.4 & & 0.070 (9) & 0.98 (18) & -18.1 & 0.073 (16) & 8.8 \\
05490+2658 & 0.042 (8) & 1.6 (3) & 0.6 & 0.07 (2) & 8.4 & & 0.044 (9) & 1.2 (2) & 1.1 & 0.057 (16) & 8.5 \\
05553+1631 & 0.042 (7) & 2.4 (4) & 4.7 & 0.11 (3) & 7.7 & & 0.042 (7) & 1.1 (2) & 6.2 & 0.047 (13) & 7.2 \\
18437-0216 & $< 0.03$ ($3\sigma$) & ... & ... & ... & 10.4 & & $< 0.03$ ($3\sigma$) & ... & ... & ... & 10.2 \\
18445-0222 & 0.126 (6) & 3.07 (17) & 87.1 & 0.41 (3) & 8.1 & & $< 0.03$ ($3\sigma$) & ... & ... & ... & 7.5 \\
18447-0229 & 0.087 (11) & 1.5 (2) & 102.3 & 0.14 (3) & 10.5 & & $< 0.03$ ($3\sigma$) & ... & ... & ... & 10.4 \\
18454-0158 & $< 0.03$ ($3\sigma$) & ... & ... & ... & 9.5 & & $< 0.03$ ($3\sigma$) & ... & ... & ... & 8.6 \\
19035+0641 & 0.172 (10) & 3.6 (2) & 31.3 & 0.65 (6) & 10.4 & & $< 0.03$ ($3\sigma$) & ... & ... & ... & 7.6 \\
19074+0752 & 0.085 (9) & 2.5 (3) & 56.1 & 0.22 (4) & 9.0 & & $< 0.03$ ($3\sigma$) & ... & ... & ... & 11.4 \\
19175+1357 & 0.064 (10) & 1.6 (2) & 14.0 & 0.11 (2) & 9.7 & & $< 0.03$ ($3\sigma$) & ... & ... & ... & 8.2 \\
19217+1651 & 0.344 (13) & 3.02 (10) & 3.3 & 1.10 (6) & 13.3 & & 0.068 (10) & 1.02 (19) & 2.9 & 0.074 (17) & 9.5 \\
19220+1432 & 0.190 (8) & 3.75 (18) & 68.9 & 0.76 (5) & 10.2 & & 0.060 (9) & 1.17 (19) & 69.5 & 0.074 (16) & 8.3 \\
19266+1745 & 0.238 (8) & 2.47 (10) & 4.9 & 0.63 (3) & 10.7 & & $< 0.03$ ($3\sigma$) & ... & ... & ... & 8.4 \\
19282+1814 & 0.046 (11) & 2.1 (5) & 23.7 & 0.10 (4) & 12.7 & & $< 0.03$ ($3\sigma$) & ... & ... & ... & 8.6 \\
19403+2258 & $< 0.03$ ($3\sigma$) & ... & ... & ... & 9.6 & & $< 0.03$ ($3\sigma$) & ... & ... & ... & 8.6 \\
19410+2336 & 0.793 (12) & 1.87 (3) & 22.5 & 1.58 (4) & 12.6 & & 0.079 (9) & 2.0 (2) & 22.3 & 0.16 (2) & 9.6 \\
19411+2306 & 0.216 (11) & 1.69 (10) & 28.9 & 0.39 (3) & 9.4 & & $< 0.03$ ($3\sigma$) & ... & ... & ... & 8.6 \\
19413+2332 & 0.136 (10) & 1.77 (15) & 20.1 & 0.26 (3) & 9.7 & & 0.028 (8) & 1.5 (4) & 19.8 & 0.045 (19) & 8.3 \\
19471+2641 & $< 0.03$ ($3\sigma$) & ... & ... & ... & 9.2 & & $< 0.03$ ($3\sigma$) & ... & ... & ... & 9.6 \\
20051+3435 & 0.146 (10) & 1.79 (14) & 11.9 & 0.28 (3) & 10.2 & & $< 0.03$ ($3\sigma$) & ... & ... & ... & 9.8 \\
20081+2720 & $< 0.03$ ($3\sigma$) & ... & ... & ... & 8.1 & & $< 0.03$ ($3\sigma$) & ... & ... & ... & 9.3 \\
20126+4104 & 0.566 (16) & 1.94 (6) & -4.0 & 1.17 (5) & 14.6 & & 0.099 (10) & 2.0 (2) & -4.5 & 0.21 (3) & 10.5 \\
20205+3948 & 0.036 (10) & 1.2 (3) & -1.8 & 0.05 (2) & 10.1 & & $< 0.03$ ($3\sigma$) & ... & ... & ... & 9.5 \\
20216+4107 & 0.225 (11) & 1.62 (9) & -1.3 & 0.39 (3) & 8.0 & & 0.052 (9) & 1.3 (2) & -3.6 & 0.073 (17) & 8.5 \\
20293+3952 & 0.304(10) & 2.32 (9) & 5.9 & 0.75 (4) & 10.0 & & 0.042 (7) & 2.1 (4) & 5.8 & 0.09 (2) & 7.9 \\
20319+3958 & 0.092 (9) & 1.24 (13) & 8.2 & 0.12 (2) & 8.9 & & $< 0.03$ ($3\sigma$) & ... & ... & ... & 9.2 \\
20332+4124 & 0.164 (11) & 2.6 (2) & -2.9 & 0.46 (5) & 13.9 & & 0.039 (9) & 1.3 (3) & -3.6 & 0.054 (19) & 9.4 \\
20343+4129 & 0.313 (16) & 1.98 (12) & 11.0 & 0.66 (5) & 13.0 & & 0.037 (10) & 1.5 (4) & 10.4 & 0.06 (2) & 9.7 \\
22134+5834 & 0.823 (15) & 1.48 (3) & -18.3 & 1.29 (4) & 13.2 & & 0.144 (10) & 1.25 (9) & -18.3 & 0.19 (2) & 10.2 \\
22551+6221 & $< 0.04$ ($3\sigma$) & ... & ... & ... & 13.7 & & $< 0.03$ ($3\sigma$) & ... & ... & ... & 9.3 \\
22570+5912 & 0.129 (13) & 2.0 (2) & -45.0 & 0.27 (4) & 13.0 & & $< 0.03$ ($3\sigma$) & ... & ... & ... & 9.0 \\
23033+5951 & 0.53 (2) & 2.41 (10) & -53.7 & 1.37 (8) & 11.7 & & 0.054 (8) & 1.6 (2) & -53.3 & 0.09 (2) & 8.5 \\
23139+5939 & 0.216 (10) & 2.71 (14) & -44.8 & 0.62 (4) & 11.5 & & $< 0.03$ ($3\sigma$) & ... & ... & ... & 8.5 \\
23151+5912 & 0.059 (9) & 2.3 (4) & -54.0 & 0.15 (3) & 10.8 & & $< 0.03$ ($3\sigma$) & ... & ... & ... & 9.2 \\
23545+6508 & $< 0.03$ ($3\sigma$) & ... & ... & ... & 9.8 & & $< 0.03$ ($3\sigma$) & ... & ... & ... & 9.7 \\
\enddata
\tablecomments{The numbers in parentheses represent one standard deviation in the Gaussian fit. The errors are written in units of the last significant digit.}
\tablenotetext{a}{These values are not corrected for instrumental velocity resolution.}
\tablenotetext{b}{The error of $V_{\rm {LSR}}$ is commonly 0.8 km s$^{-1}$, corresponding to the velocity resolution of the final spectra (Section \ref{sec:obs}).}
\tablenotetext{c}{The rms noises in emission-free region.}
\end{deluxetable}

\begin{figure}
\figurenum{1}
\plotone{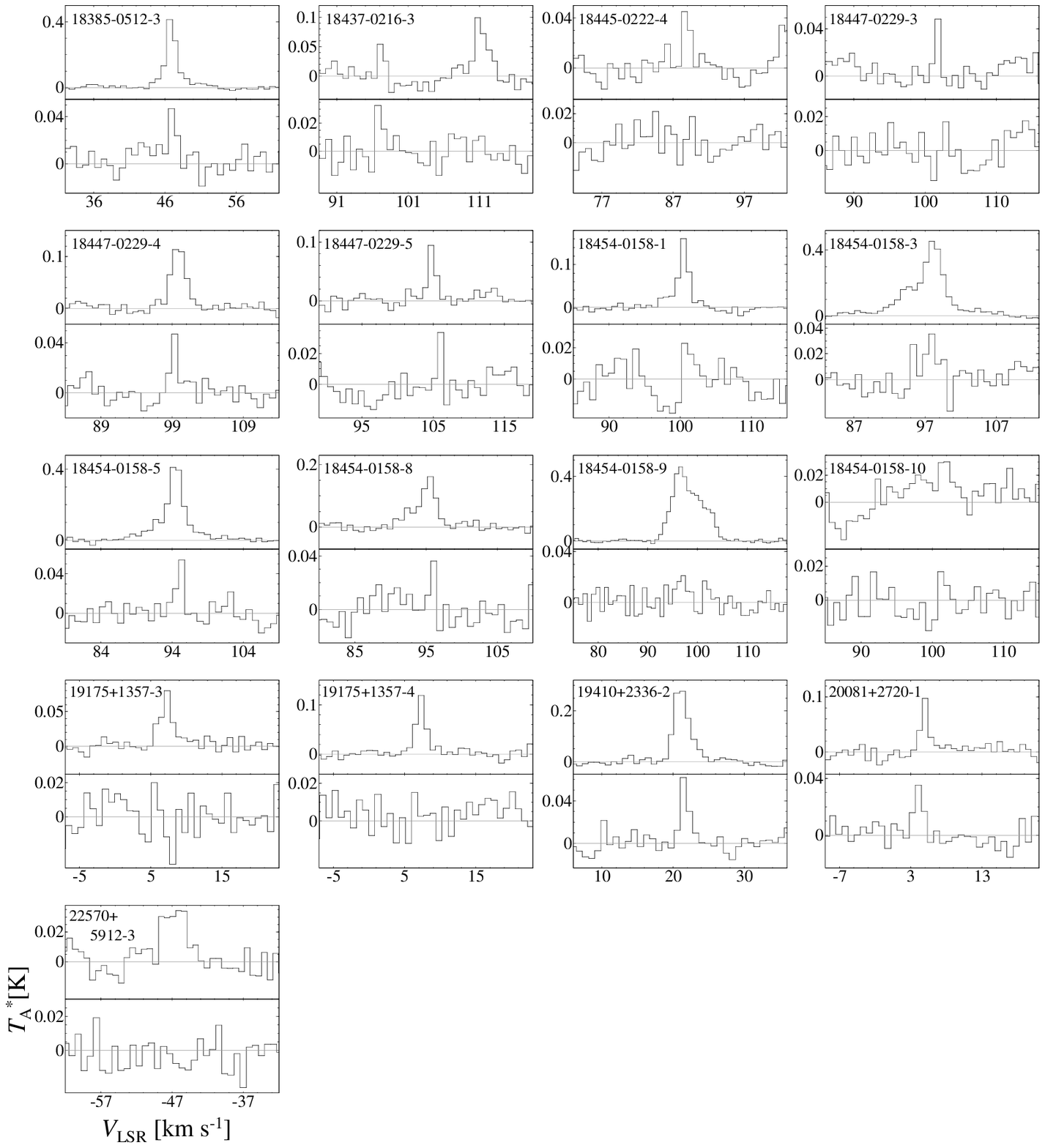}
\caption{Spectra of HC$_{3}$N (upper) and HC$_{5}$N (lower) in HMSCs.\label{fig:f1}}
\end{figure}

\begin{figure}
\figurenum{2}
\plotone{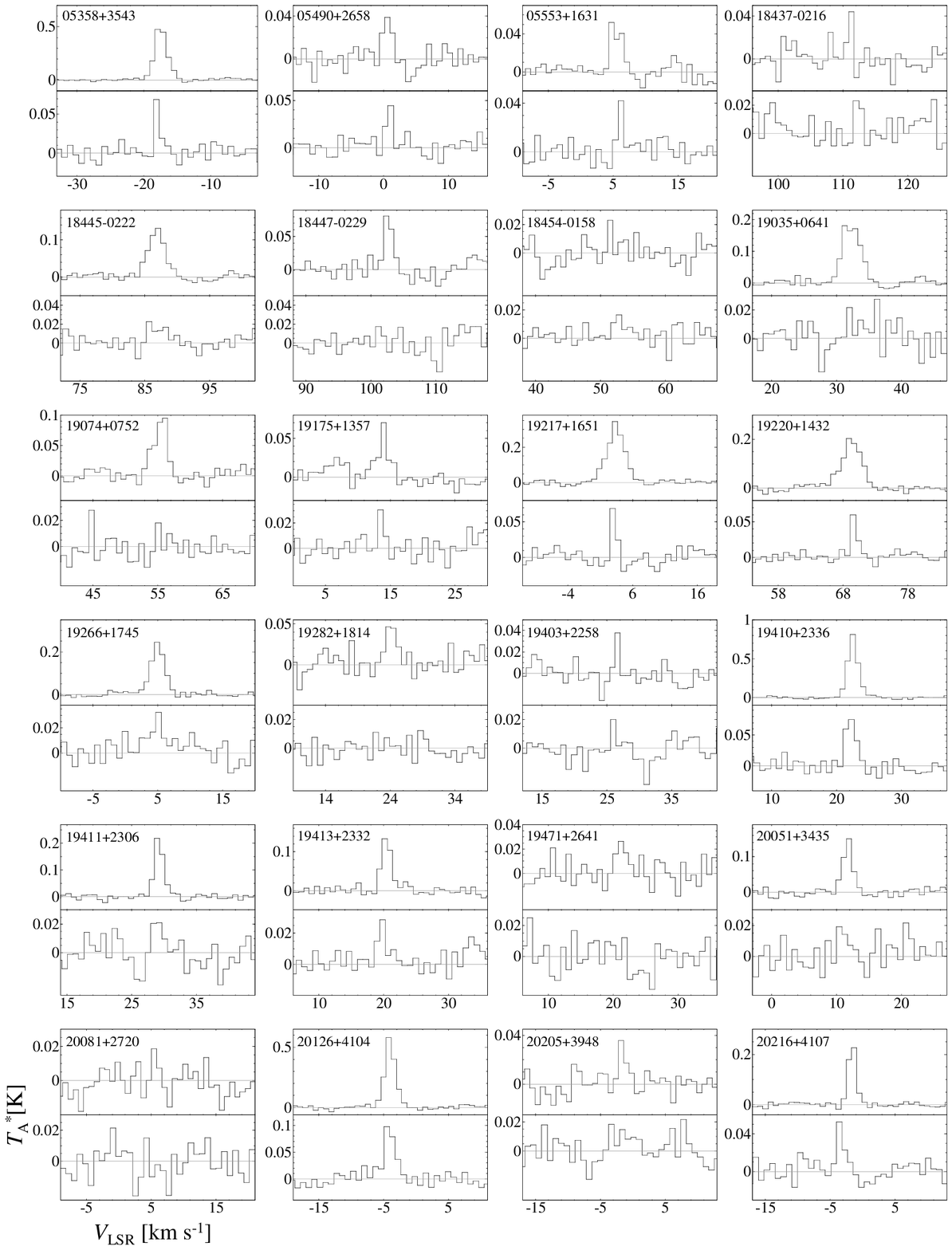}
\caption{Spectra of HC$_{3}$N (upper) and HC$_{5}$N (lower) in HMPOs.\label{fig:f2}}
\end{figure}

\begin{figure}
\figurenum{3}
\plotone{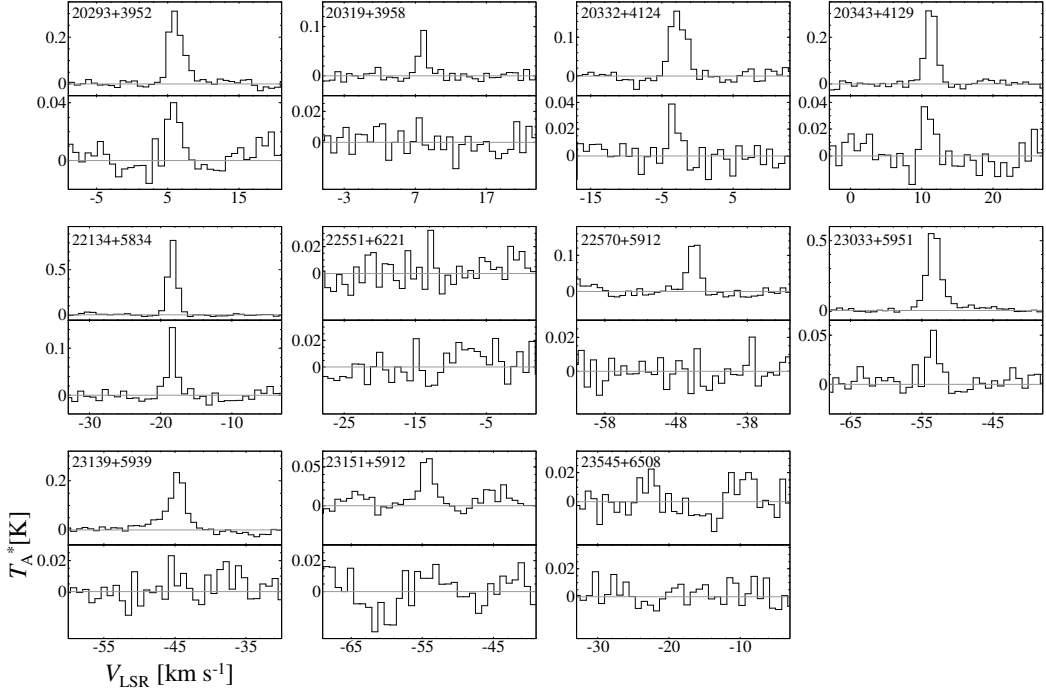}
\caption{Continued.\label{fig:f3}}
\end{figure}

\subsection{Analysis}

The lower state excitation energies of the observed HC$_{3}$N and HC$_{5}$N lines are 4.4 K and 15.3 K, respectively.
The critical densities of these lines are $\sim 10^{4}$ cm$^{-3}$.
The typical gas density of the observed sources is $10^{5} - 10^{6}$ cm$^{-3}$ \citep{2002ApJ...566..945B}.
Therefore, the local thermodynamic equilibrium (LTE) assumption is reasonable.
We then derived the column densities of HC$_{3}$N and HC$_{5}$N, assuming the LTE.
We used the following formulae: 

\begin{equation} \label{tau}
\tau = - {\mathrm {ln}} \left[1- \frac{T_{\rm A}^{*}}{f\eta_{\rm {B}}\left\{J(T_{\rm {ex}}) - J(T_{\rm {bg}}) \right\}} \right],  
\end{equation}
where
\begin{equation} \label{tem}
J(T) = \frac{h\nu}{k}\Bigl\{\exp\Bigl(\frac{h\nu}{kT}\Bigr) -1\Bigr\} ^{-1},
\end{equation}  
and
\begin{equation} \label{col}
N = \tau \frac{3h\Delta v}{8\pi ^3}\sqrt{\frac{\pi}{4\mathrm {ln}2}}Q\frac{1}{\mu ^2}\frac{1}{J_{\rm {lower}}+1}\exp\Bigl(\frac{E_{\rm {lower}}}{kT_{\rm {ex}}}\Bigr)\Bigl\{1-\exp\Bigl(-\frac{h\nu }{kT_{\rm {ex}}}\Bigr)\Bigr\} ^{-1}.
\end{equation} 
In Equation (\ref{tau}), $T_{\rm A}^{*}$, $f$, $\eta_{\rm {B}}$, and $\tau$ denote the antenna temperature, the beam filling factor, the main beam efficiency, and the optical depth, respectively.
We used 0.72 for the main beam efficiency (Section \ref{sec:obs}).
We used 1 for the beam filling factor, because we do not know the exact spatial distributions of HC$_{3}$N and HC$_{5}$N.
Therefore, the derived column densities are the beam-averaged values.
$T_{\rm {ex}}$ and $T_{\rm_{bg}}$ are the excitation temperature and the cosmic microwave background temperature (2.7 K), respectively.
The brightness temperature of dust emission is estimated below 0.04 K at the observed frequency band, which we neglect.
We used the rotational temperatures of NH$_{3}$ summarized in \citet{2002ApJ...566..931S, 2005ApJ...634L..57S} as the excitation temperature in our calculations.
In case that we cannot obtain the rotational temperatures of NH$_{3}$ for each source, we used the mean values of its rotational temperature, 15.3 K for HMSCs and 22.5 K for HMPOs \citep{2005ApJ...634L..57S}, respectively. 
We investigated the relationship between the derived column densities and the used excitation temperatures, but we did not recognize any tendency.
Therefore, the analysis using the mean values of the rotational temperature does not significantly affect the results.
$J$($T$)  in Equation (\ref{tem}) is the Planck function.
In Equation (\ref{col}), $N$ denotes the column density, $\Delta v$ the line widths (FWHM), $Q$ the partition function ($Q = \frac{kT_{\rm {ex}}}{hB}$; $B$ is the rotational constant; 4549.059 MHz for HC$_{3}$N and 1331.3327 MHz for HC$_{5}$N, which were taken from Splatalogue database for astronomical spectroscopy\footnote{http://www.cv.nrao.edu/php/splat/advanced.php}), $\mu$ the permanent electric dipole moment, and $E_{\rm {lower}}$ the energy of the lower rotational energy level.
The permanent electric dipole moments of HC$_{3}$N and HC$_{5}$N are 3.7317 D \citep{1985JChPh...82...1702D} and 4.33 D \citep{1976JMoSp...62...175A}, respectively.

The column densities of HC$_{3}$N and HC$_{5}$N in HMSCs and HMPOs are summarized in Tables \ref{tab:tab3} and \ref{tab:tab4}.
We also derived their fractional abundances ($X$($a$) = $N$($a$)/$N_{\rm {gas}}$; $N_{\rm {gas}}$ = $N_{\rm {H_{2}}}$) using the values of $N_{\rm {gas}}$ summarized in Table 3 in \citet{2002ApJ...566..945B} and in Table 2 in \citet{2007ApJ...668..348B}.

\begin{figure}
\figurenum{5}
\plotone{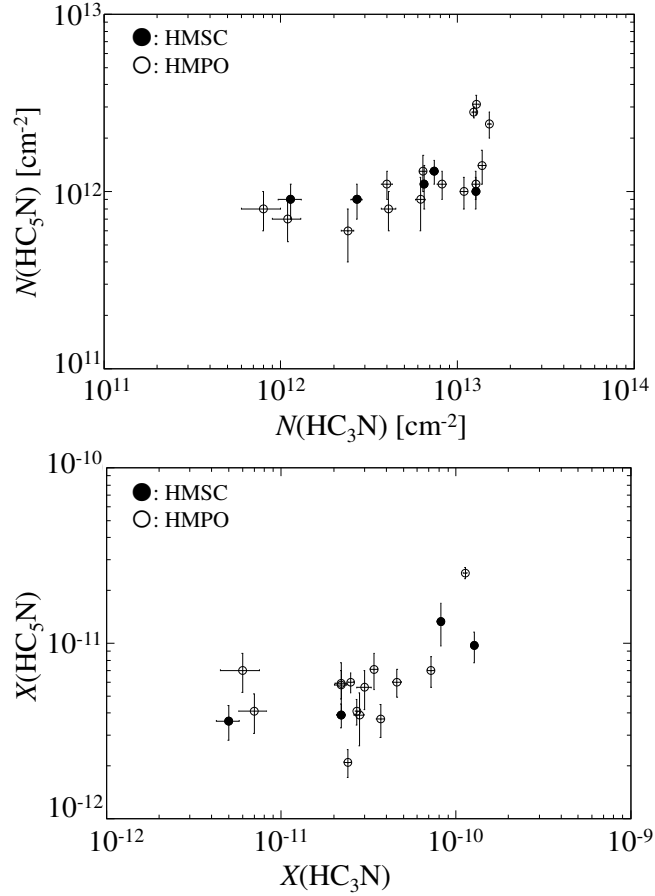}
\caption{Plot of the column density (upper) and fractional abundance (lower) of HC$_{3}$N vs. HC$_{5}$N. The bars indicate $1\sigma$ errors derived from Gaussian fitting.\label{fig:f8}}
\end{figure}

Figure \ref{fig:f8} shows the relationships of the column density and fractional abundance between HC$_{3}$N and HC$_{5}$N.
We expected a positive correlation between them, as in the low-mass star-forming regions \citep{1992ApJ...392..551S}.
In order to confirm the correlation between HC$_{3}$N and HC$_{5}$N, we conducted the Kendall's rank correlation statics.
The Kendall's tau correlation coefficients ($\tau$) for the column density and fractional abundance are derived to be $+0.61$ and $+0.43$, respectively.
The probabilities that there is no correlation between HC$_{3}$N and HC$_{5}$N are 0.04\% for the column density and 1.5\% for the fractional abundance, respectively.
Hence, we conclude that there is a positive correlation between HC$_{3}$N and HC$_{5}$N.
\floattable
\begin{deluxetable}{lccccccc}
\tabletypesize{\scriptsize}
\tablecaption{Column Densities and Fractional Abundances of HC$_{3}$N and HC$_{5}$N in HMSCs\label{tab:tab3}}
\tablewidth{0pt}
\tablehead{
\colhead{Source} & \colhead{$N$(HC$_{3}$N)} & \colhead{$X$(HC$_{3}$N)} &  \colhead{$N$(HC$_{5}$N)} & \colhead{$X$(HC$_{5}$N)} & \colhead{$S_{1.2{\rm {mm}}}$\tablenotemark{a}} & \colhead{$N_{\rm {gas}}$\tablenotemark{a}} & \colhead{$T_{\rm {rot}}$(NH$_{3}$)\tablenotemark{b}} \\
\colhead{} & \colhead{($\times 10^{12}$ cm$^{-2}$)} & \colhead{($\times 10^{-11}$)} & \colhead{($\times 10^{11}$ cm$^{-2}$)} & \colhead{($\times 10^{-12}$)} & \colhead{(mJy)} & \colhead{($\times 10^{23}$ cm$^{-2}$)} & \colhead{(K)}
}
\startdata
18385-0512-3 & 6.5 (3) & 8.2 (4) & 11 (3) & 13 (4) & 77 & 0.8 & 14.4 \\
18437-0216-3 & 1.8 (3) & 1.8 (3) & ...& ...& 96 & 1.0 & 15.9 \\
18445-0222-4 & 0.54 (17) & ... & ... & ... & ... & ... & {\it {15.3}} \\
18447-0229-4 & 2.7 (2) & ... & 9 (2) & ... & ... & ... & {\it {15.3}} \\
18447-0229-5 & 1.12 (15) & ... & ... & ... & ... & ... & {\it {15.3}} \\
18454-0158-1 & 2.3 (2) & 1.3 (1) & ... & ...& 178 & 1.8 & {\it {15.3}} \\
18454-0158-3 & 17.5 (7) & 8.7 (3) & ... & ... & 204 & 2.0 & {\it {15.3}} \\
18454-0158-5 & 12.7 (6) & 12.7 (6) & 10 (2) & 10 (2) & 102 & 1.0 & 17.6\\
18454-0158-8 & 3.6 (4) & ... & ... & ... & ... & ... & {\it {15.3}} \\
19175+1357-3 & 1.5 (2) & 2.9 (4) & ... & ... & 53 & 0.5 & 15.6 \\
19175+1357-4 & 1.59 (17) & 1.8 (2) & ... & ... & 87 & 0.9 & 13.2 \\
19410+2336-2 & 7.4 (4) & 2.2 (1) & 13 (2) & 3.9 (6) & 343 & 3.4 & 18.3 \\
20081+2720-1 & 1.14 (17) & 0.5 (1) & 9 (2) & 3.6 (8) & 180 & 2.4 & {\it {15.3}} \\
22570+5912-3 & 1.3 (3) & 1.5 (3) & ... & ... & 88 & 0.9 & 16.2 \\
\enddata
\tablecomments{The numbers in parentheses represent one standard deviation.}
\tablenotetext{a}{The values are taken from \citet{2007ApJ...668..348B}.}
\tablenotetext{b}{The values are taken from \citet{2005ApJ...634L..57S}. The italic letter indicates the average value of HMSCs (15.3 K).}
\end{deluxetable}
\floattable
\begin{deluxetable}{lccccccc}
\tabletypesize{\scriptsize}
\tablecaption{Column Densities and Fractional Abundances of HC$_{3}$N and HC$_{5}$N in HMPOs\label{tab:tab4}}
\tablewidth{0pt}
\tablehead{
\colhead{Source} & \colhead{$N$(HC$_{3}$N)} & \colhead{$X$(HC$_{3}$N)} &  \colhead{$N$(HC$_{5}$N)} & \colhead{$X$(HC$_{5}$N)} & \colhead{$S_{1.2{\rm {mm}}}$\tablenotemark{a}} & \colhead{$N_{\rm {gas}}$\tablenotemark{a}} & \colhead{$T_{\rm {rot}}$(NH$_{3}$)\tablenotemark{b}} \\
\colhead{} & \colhead{($\times 10^{12}$ cm$^{-2}$)} & \colhead{($\times 10^{-11}$)} & \colhead{($\times 10^{11}$ cm$^{-2}$)} & \colhead{($\times 10^{-12}$)} & \colhead{(mJy)} & \colhead{($\times 10^{23}$ cm$^{-2}$)} & \colhead{(K)}
}
\startdata
05358+3543 & 10.9 (5) & 7.2 (3) & 10 (2) & 7 (1) & {\it {102}} & {\it {1.5}} & 18 \\
05490+2658 & 0.8 (2) & 0.6 (2) & 8 (2) & 7 (2) & 176 & 1.2 & {\it {22.5}} \\
05553+1631 & 1.1 (2) & 0.7 (1) & 7.0 (1.8) & 4 (1) & 317 & 1.7 & {\it {22.5}} \\
18445-0222 & 4.2 (3) & 2.3 (2) & ... & ... & 257 & 1.8 & 21 \\
18447-0229 & 1.2 (2) & 1.3 (2) & ... & ... & {\it {98}} & {\it {0.9}} & 15 \\
19035+0641 & 6.7 (6) & 3.5 (3) & ... & ... & 312 & 1.9 & 21 \\
19074+0752 & 1.9 (3) & 2.1 (3) & ... & ... & 129 & 0.9 & 16 \\
19175+1357 & 1.2 (2) & 0.9 (2) & ... & ... & 141 & 1.3 & {\it {22.5}} \\
19217+1651 & 12.7 (6) & 2.4 (1) &11 (2) & 2.1 (4) & 640 & 5.3 & 25 \\
19220+1432 & 8.2 (5) & 4.6 (3) & 11 (2) & 6 (1) & 256 & 1.8 & 23 \\
19266+1745 & 6.7 (3) & 2.0 (1) & ... & ... & {\it {323}} & {\it {3.3}} & {\it {22.5}} \\
19282+1814 & 1.1 (4) & 0.5 (2) & ... & ... & {\it {273}} & {\it {2.4}} & {\it {22.5}} \\
19410+2336 & 15.2 (3) & 2.7 (1) & 24 (4) & 4.1 (7) & 849 & 5.7 & 18 \\
19411+2306 & 3.2 (2) & 1.9 (1) & ... & ... & 222 & 1.7 & 14 \\
19413+2332 & 2.4 (2) & 2.2 (2) & 6 (2) & 6 (2) & 139 & 1.1 & 18 \\
20051+3435 & 3.0 (3) & 3.0 (3) & ... & ... & 167 & 1.0 & {\it {22.5}} \\
20126+4104 & 12.8 (5) & 2.5 (1) & 31 (4) & 6.0 (8) & 1087 & 5.2 & 23 \\
20205+3948 & 0.5 (2) & 0.7 (3) & ... & ... & {\it {104}} & {\it {0.7}} & {\it {22.5}} \\
20216+4107 & 4.0 (3) & 2.2 (2) & 11 (2) & 6 (1)  & 264 &1.8 & 21 \\
20293+3952 & 6.4 (3) & 3.4 (2) & 13 (3) & 7 (2) & 354 & 1.9 & 15 \\
20319+3958 & 1.30 (18) & 1.4 (2) & ... & ... & 214 & 0.9 & {\it {22.5}} \\
20332+4124 & 4.1 (4) & 3.0 (3) & 8 (2) & 6 (1) & {\it {265}} & {\it {1.4}} & 17 \\
20343+4129 & 6.2 (4) & 2.8 (2) & 9 (3) & 4 (1) & {\it {313}} & {\it {2.2}} & 18 \\
22134+5834 & 12.4 (3) & 11.3 (3) & 28 (2) & 25 (2) & 229 & 1.1 & 18 \\
22570+5912 & 2.9 (4) & 3.2 (4) & ... & ... &157 & 0.9 & {\it {22.5}} \\
23033+5951 & 13.8 (7) & 3.7 (2) & 14 (3) & 3.7 (8) & 631 & 3.7 & 20 \\
23139+5939 & 6.7 (4) & 1.7 (1) & ... & ... & 530 & 4.0 & {\it {22.5}} \\
23151+5912 & 1.6 (3) & 0.9 (2) & ... & ... & 406 & 1.8 & {\it {22.5}} \\
\enddata
\tablecomments{The numbers in parentheses represent one standard deviation.}
\tablenotetext{a}{The values are taken from \citet{2002ApJ...566..945B}. The italic letter indicates that we do not use in the statistical analyses due to core ambiguity.}
\tablenotetext{b}{The values are taken from \citet{2002ApJ...566..931S}. The italic letter indicates the average value of HMPOs (22.5 K).}
\end{deluxetable}
\\
\section{Discussion}

We mainly treat the results of HC$_{3}$N, because the sample size of the HC$_{5}$N-detected sources is small.
In the statistical analyses, we use the results in 16 HMSCs and 27 HMPOs, adding the data from \citet{2008ApJ...678...1049S} using the Nobeyama 45 m telescope (seven HMSCs and six HMPOs, summarized in Appendix \ref{a1}).
We exclude seven HMPOs (Table \ref{tab:tab4}), because our observing positions are off from the 1.2 mm continuum cores.
We also exclude HMSC 20081+2720-1, because the observed position looks like a HMPO core and the position is not at the exact 1.2 mm continuum core.

\subsection{The chemical evolution of HC$_{3}$N from HMSCs to HMPOs} \label{sec:d2}

\begin{figure}
\figurenum{6}
\plotone{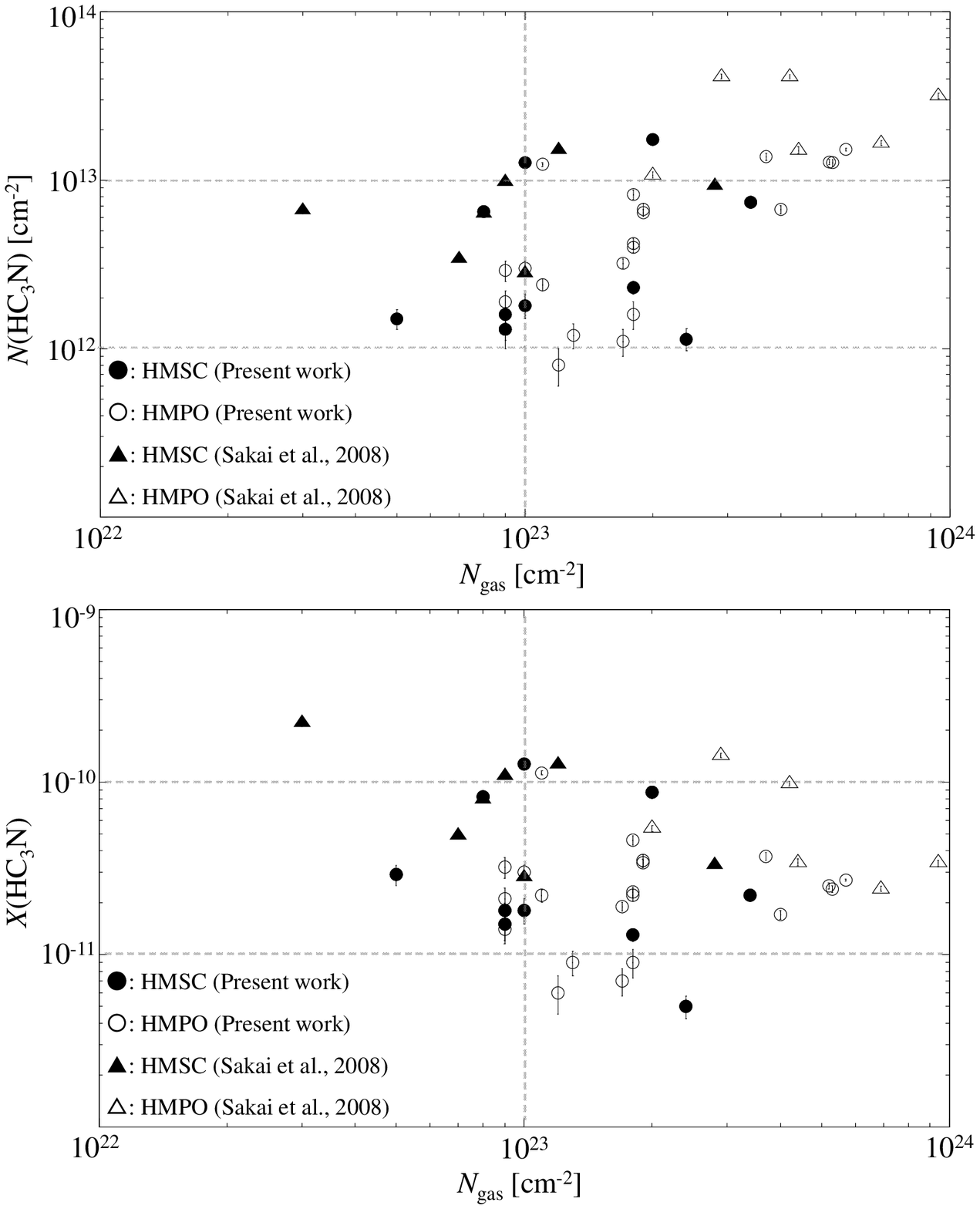}
\caption{Plots of the column density of HC$_{3}$N ($N$(HC$_{3}$N)) vs. H$_{2}$ ($N_{gas}$) (upper) and the fractional abundance of HC$_{3}$N ($X$(HC$_{3}$N)) vs. H$_{2}$ ($N_{gas}$) (lower). The bars indicate $1\sigma$ errors.\label{fig:f4}}
\end{figure}

\begin{figure}
\figurenum{7}
\plotone{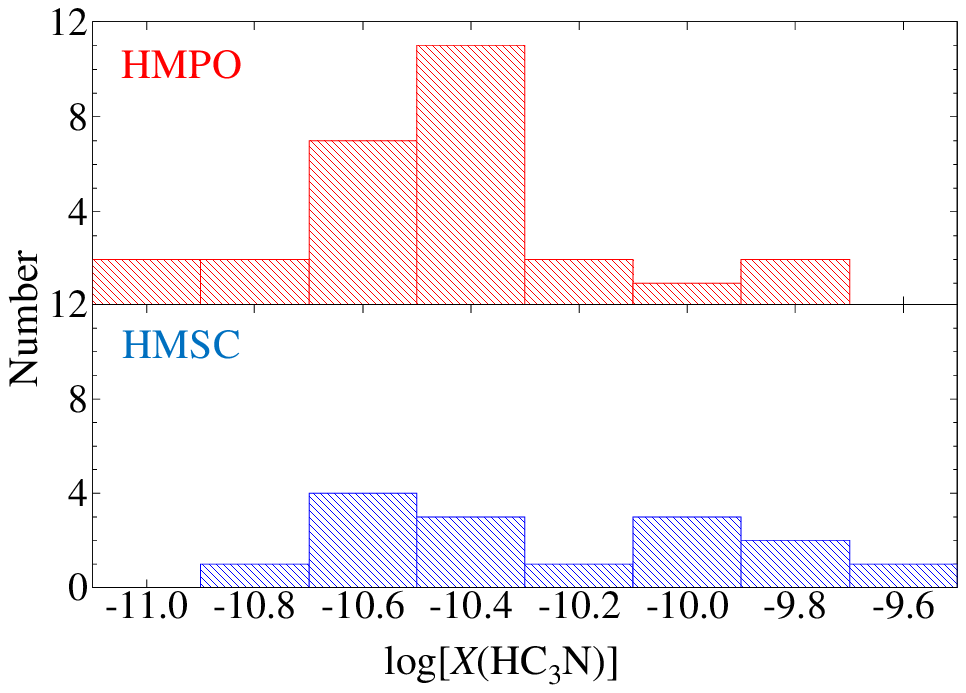}
\caption{Histogram of the fractional abundance of HC$_{3}$N ($X$(HC$_{3}$N)) in HMSCs (blue) and HMPOs (red).\label{fig:f7}}
\end{figure}

\citet{2008ApJ...678...1049S} found the positive correlation between the integrated intensity of HC$_{3}$N and  the peak flux of the 1.2 mm continuum.
From the results, they suggested that HC$_{3}$N emission comes from a dense gas region.
Figure \ref{fig:f4} shows the relationship between the gas column density $N_{\rm {gas}}$ and the column density and fractional abundance of HC$_{3}$N, $N$(HC$_{3}$N) and $X$(HC$_{3}$N), including sources from \citet{2008ApJ...678...1049S}.
In order to quantify the correlation, we conduct the Kendall's rank correlation statics between $N$(HC$_{3}$N) and $N_{gas}$.
The Kendall's tau correlation coefficients ($\tau$) are derived to be $+0.16$ and $+0.63$ for HMSCs and HMPOs, respectively.
The probabilities that there is no correlation between $N$(HC$_{3}$N) and $N_{gas}$ are 38.4\% for HMSCs and $5 \times 10^{-4}$\% for HMPOs, respectively.
Thus, for HMSCs, $N$(HC$_{3}$N) is independent of $N_{gas}$, which may imply the chemical diversity.
The positive correlation between $N$(HC$_{3}$N) and $N_{\rm {gas}}$ in HMPOs suggests that HC$_{3}$N exists in the dense gas.

The fractional abundance of HC$_{3}$N, $X$(HC$_{3}$N), shows a range of $\sim 10^{-11} - 10^{-10}$ independently of the gas column density (the bottom figure of Figure \ref{fig:f4}).
Figure \ref{fig:f7} shows the histograms of the fractional abundance of HC$_{3}$N, $X$(HC$_{3}$N), in HMSCs and HMPOs.
We investigate the distribution of $X$(HC$_{3}$N) in HMSCs and HMPOs using the Kolmogorov-Smirnov test (K-S test).
The possibility that the fractional abundances of HC$_{3}$N in HMSCs and HMPOs originate from the same parent population is 17\%.
The distributions of the fractional abundance of HC$_{3}$N may be different between HMSCs and HMPOs.
The mean values of $X$(HC$_{3}$N) in HMSCs and HMPOs are derived to be ($6.6 \pm 0.8$)$\times 10^{-11}$ and ($3.6 \pm 0.5$)$\times 10^{-11}$, respectively.
While the trend is weak with a small difference, $X$(HC$_{3}$N) decreases from HMSCs to HMPOs.
This may imply that the higher density is required for the existence of HC$_{3}$N in HMPOs, as we discuss in Section \ref{sec:d3}.

We also conduct the same analyses about the column density.
From the K-S test, the possibility that the column densities of HC$_{3}$N in HMSCs and HMPOs originate from the same parent population is 43\%. 
The average values of $N$(HC$_{3}$N) are derived to be ($5.7 \pm 0.7$)$\times 10^{12}$ and ($1.03 \pm 0.12$)$\times 10^{13}$ cm$^{-2}$ in HMSCs and HMPOs, respectively. 
Again, while the difference is small, the column density of HC$_{3}$N seems to increase from HMSCs to HMPOs.
Although the results are marginal, this trend with evolution is opposite to that found in the low-mass star-forming regions \citep{1992ApJ...392..551S}.

Our statistical analyses show that the column density of HC$_{3}$N is not clearly correlated with the gas column density $N_{gas}$ in HMSCs, while there is a positive correlation between them in HMPOs.
On the other hand, the fractional abundance of HC$_{3}$N decreases from HMSCs to HMPOs.
From the chemical point of view, HC$_{3}$N appears to be newly formed from chemical species evaporated from grain mantles such as CH$_{4}$ and/or C$_{2}$H$_{2}$ in the warm gas in HMPOs, as discussed later in detail.
Such a formation mechanism is unlikely to occur in HMSCs without any heating sources.
Instead, HC$_{3}$N is considered to be formed by the gas-phase reactions in HMSCs as in low-mass starless cores.
From the non-LTE analysis using the RADEX code \citep{2007A&A...468..627V}, given the same HC$_{3}$N column density, the line intensity of the observed transition of HC$_{3}$N is insensitive to the changes of the temperature ($T_{\rm {kin}} = 15-25$ K) and becomes slightly lower with the increase in the gas density ($n$(H$_{2}$)$=10^{5}-10^{6}$ cm$^{-3}$).
The gas density increases from HMSCs ($10^{5}$ cm$^{-3}$) to HMPOs ($10^{6}$ cm$^{-3}$) \citep{2002ApJ...566..945B}, but the observed line intensities in HMPOs are not low compared to those in HMSCs.
Hence, the physical conditions cannot explain the increase in $N$(HC$_{3}$N) from HMSCs to HMPOs.

We derive the errors of the average values of the column density and fractional abundance taking both the Gaussian fitting errors and the absolute calibration error of 10\% (Section \ref{sec:obs}) into consideration.
The conclusions do not change due to these errors.

The increase in the column density of HC$_{3}$N from HMSCs to HMPOs suggests the formation of HC$_{3}$N in HMPOs. 
HC$_{3}$N can be formed from chemical species evaporated from grain mantles in the warm gas.
There are two possible key species evaporated from grain mantles; CH$_{4}$ and C$_{2}$H$_{2}$.
In the low-mass star-forming cores, CH$_{4}$ evaporated from grain mantles produces carbon chains, and such chemistry was named warm carbon chain chemistry \citep[WCCC,][]{2013ChRv...113...8981}.
\citet{2009MNRAS.394..221C} suggested that C$_{2}$H$_{2}$ could efficiently form cyanopolyynes in hot cores.
The sublimation temperatures of CH$_{4}$ and C$_{2}$H$_{2}$ are approximately 25 K and 50 K, respectively \citep{1983A&A...122...171Y}.
As the dust temperatures in HMPOs \citep{2002ApJ...566..931S} are much higher than these sublimation temperatures, both species can evaporate into the gas phase and be potential parent species for HC$_{3}$N.
In addition, the higher density and temperature accelerate the chemical reaction rates.
In order to result in an increased column density compared to the HMSC stage, the formation of HC$_{3}$N appears to proceed more efficiently than its destruction in the high-mass star-forming regions.

The contradiction between the marginal increase in $N$(HC$_{3}$N) and the slight decrease in $X$(HC$_{3}$N) is caused by the increase in $N_{gas}$ as seen in the x-axis distribution of Figure \ref{fig:f4}.
Its average values in HMSCs and HMPOs are derived to be $1.3 \times 10^{24}$ cm$^{-2}$ and $2.8 \times 10^{24}$ cm$^{-2}$, respectively.
From the K-S test, the possibility that the gas column densities in HMSCs and HMPOs originate from the same parent population is 1.1\%.
This may imply that the higher gas density is needed for existence of HC$_{3}$N in HMPOs, which is consistent with the suggestion from the positive correlation between $N$(HC$_{3}$N) and $N_{gas}$ in HMPOs.
This may be due to the fact that there is significant UV radiation in HMPOs and the higher densities provide better shielding.

We also investigate the chemical evolution locally using the data of HMSCs and HMPOs in the same fields, thus likely to be physically located in the same regions.
The sample size to derive fractional abundances of HC$_{3}$N in both HMSCs and HMPOs is deficient to test statistically.
We then investigate in individual regions and found that the tendencies are different for each region.
For example, in 19175+1357, the values of $X$(HC$_{3}$N) in HMSCs ($2.9 \times 10^{-11}$ and $1.8 \times 10^{-11}$) are higher than that in HMPO ($9 \times 10^{-12}$).
In 22570+5912, on the other hand, the value of $X$(HC$_{3}$N) in HMSC ($1.5 \times 10^{-11}$) is lower than that in HMPO ($3.2 \times 10^{-11}$).
We cannot achieve conclusions.

\subsection{The relationship between the column density of HC$_{3}$N and the luminosity-to-mass ratio} \label{sec:d3}

\begin{figure}
\figurenum{8}
\plotone{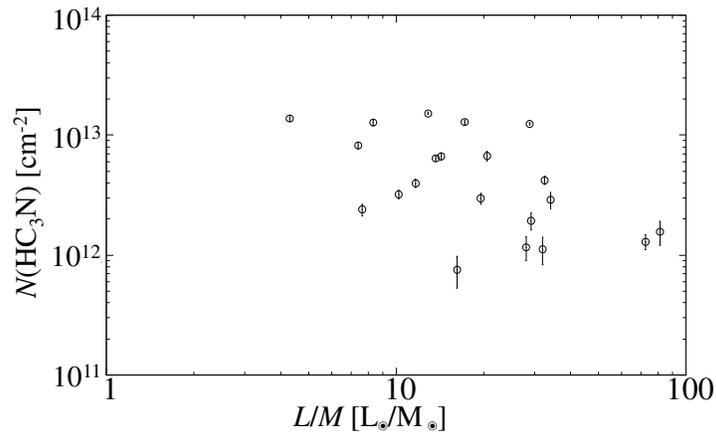}
\caption{Plot of the column density of HC$_{3}$N ($N$(HC$_{3}$N)) vs. the luminosity-to-mass ratio ($L/M$). We take the values of luminosity and mass from \citet{2002ApJ...566..931S}. The bars indicate $1\sigma$ errors. \label{fig:f5}}
\end{figure}

We investigate the relationship between the column density of HC$_{3}$N, $N$(HC$_{3}$N), and the luminosity-to-mass ratio, $L/M$, which is a physical evolutional indicator \citep{2002ApJ...566..931S}, in HMPOs as shown in Figure \ref{fig:f5}.
We take the values of luminosity and mass from \citet{2002ApJ...566..931S}.
We only use results from our HMPO sample observations, excluding the seven sources where our observing positions are off from the 1.2 mm continuum cores.

In order to investigate the relationship between the physical evolution and chemistry, we conduct the Kendall's rank correlation test between $N$(HC$_{3}$N) and the $L/M$ ratio.
The Kendall's tau correlation coefficient ($\tau$) is derived to be $-0.34$.
The probability that there is no correlation between $N$(HC$_{3}$N) and the $L/M$ ratio is derived to be 3.1\%.
The column density of HC$_{3}$N tends to decrease as the progress of the massive star formation.
The tendency suggests that HC$_{3}$N is destroyed by the stellar activities such as the UV radiation.
For instance, \citet{2012ApJ...753...34J} carried out observations toward AFGL2591 high-mass star-forming region with the Submillimeter Array (SMA) and found that the spatial distribution of HC$_{3}$N has the double-peaked structure circumventing the continuum peak.
This result suggests that HC$_{3}$N survives in the dense region with the higher extinction and is destroyed in diffuse regions just near the massive young stellar objects where gas is dispersed.

\subsection{Comparison of the detection rate of HC$_{5}$N between HMPOs and hot cores} \label{sec:d4}

Our survey observations include the second largest sample size of HC$_{5}$N in the high-mass star-forming regions.
The detection rate of HC$_{5}$N in HMPOs is calculated to be 50\%, excluding the HC$_{3}$N-undetected sources.
\citet{2014MNRAS.443.2252G} carried out a survey observation of HC$_{5}$N toward 79 hot cores associated with the 6.7 GHz methanol masers with the Tidbinbilla 34 m telescope.
They detected HC$_{5}$N from 35 sources, and the detection rate was $\sim 44$\%.
The detection rate in HMPOs of our observations is roughly consistent with that in hot cores of \citet{2014MNRAS.443.2252G}.

Two independent survey observations derived the similar detection rates.
These similar detection rates seem to arise from the difficulty in distinguishing between HMPOs and hot cores, because HMPOs and hot cores are overlapped \citep{2014A&A...563A..97G}.
We consider that our observational results confirm the previous survey observation conducted by \citet{2014MNRAS.443.2252G}, and the detection rate of HC$_{5}$N is independent of the two specific source samples.
These results may imply that a half of the HMPO/hot core stage sources have associated dense gas with conditions where HC$_{5}$N can form and/or survive.

\section{Conclusions}

We carried out the survey observations of HC$_{3}$N and HC$_{5}$N in the 42$-$45 GHz band toward 17 HMSCs and 35 HMPOs with the Nobeyama 45 m radio telescope.
We detected HC$_{3}$N from 15 HMSCs and 28 HMPOs, and detected HC$_{5}$N from 5 HMSCs and 14 HMPOs, respectively.
We found the positive correlations in the column density and fractional abundance between HC$_{3}$N and HC$_{5}$N, using the Kendall's rank correlation statics.
The line widths of HC$_{5}$N are narrower than those of HC$_{3}$N in most of the sources.
We found that the fractional abundance of HC$_{3}$N decreases from HMSCs to HMPOs.
On the other hand, the average value of the column density of HC$_{3}$N slightly increases, which may suggest that HC$_{3}$N is formed in the dense gas at the HMPO stage.
We also investigate the relationship between the column density of HC$_{3}$N and the luminosity-to-mass ratio, which is a physical evolutional indicator.
The column density of HC$_{3}$N tends to decrease with increasing the luminosity-to-mass ratio.
This result suggests that HC$_{3}$N is destroyed by the stellar activities such as the UV radiation from central stars.
In the case of HC$_{5}$N, the detection rate in HMPOs is similar to that in hot cores derived from the previous survey observation.




\acknowledgments

We thank the anonymous referee who gave us valuable comments and helped us improve the quality of this paper.
We would like to express our great thanks to the staff of the Nobeyama Radio Observatory.
The Nobeyama Radio Observatory is a branch of the National Astronomical Observatory of Japan, National Institutes of Natural Sciences.
K. T. appreciates Dr. Tomomi Shimoikura for helping to search the off-source positions.
The Z45 receiver is supported in part by a Granting-Aid for Science Research of Japan (24244017).



\vspace{5mm}
\facilities{Nobeyama 45 m radio telescope}
\software{Java NEWSTAR}

\appendix

\section{Cited Data Information}  \label{a1}

In Section \ref{sec:d2}, we cited some data from \citet{2008ApJ...678...1049S}.
We summarize the cited data in Table \ref{tab:taba1}.

\floattable
\begin{deluxetable}{lcccccc}
\tabletypesize{\scriptsize}
\tablecaption{Cited data \label{tab:taba1}}
\tablewidth{0pt}
\tablehead{
\colhead{Source} & \colhead{$\int T^{\ast}_{\mathrm A}dv$\tablenotemark{a}} & \colhead{$N$(HC$_{3}$N)} & \colhead{$X$(HC$_{3}$N)} & \colhead{$S_{1.2{\rm {mm}}}$\tablenotemark{b}} & \colhead{$N_{\rm {gas}}$\tablenotemark{b}} & \colhead{$T_{\rm {rot}}$(NH$_{3}$)} \\
\colhead{} & \colhead{(K km s$^{-1}$)} & \colhead{($\times 10^{12}$ cm$^{-2}$)} & \colhead{($\times 10^{-11}$)} & \colhead{(mJy)} & \colhead{($\times 10^{23}$ cm$^{-2}$)} & \colhead{(K)} 
}
\startdata
\multicolumn{6}{l}{HMPO} \\
I18102-1800 MM1 & 4.14 (9) & $41.2^{+1.4}_{-1.2}$ & $14.2^{+0.5}_{-0.4}$ & 316 & 2.9 & $17.9^{+1.2}_{-1.1}$ \\
I18151-1208 MM1 & 1.39 (9) & $15.0^{+0.8}_{-0.7}$ & $3.4^{+0.2}_{-0.2}$ & 672 & 4.4 & $20.8^{+1.8}_{-1.7}$ \\
I18182-1433 MM1 & 3.07 (9) & $31.6^{+1.3}_{-1.2}$ & $3.4^{+0.1}_{-0.2}$ & 1303 & 9.4 & $19.0^{+1.5}_{-1.4}$ \\
I18223-1243 MM1 & 1.09 (8) & $10.7^{+0.5}_{-0.4}$ & $5.4^{+0.2}_{-0.3}$ & 328 & 2.0 & $17.5^{+1.6}_{-1.5}$ \\
I18306-0835 MM1 & 1.70 (9) & $16.6^{+0.6}_{-0.5}$ & $2.4^{+0.1}_{-0.1}$ & 731 & 6.9 & $17.3^{+1.2}_{-1.1}$ \\
I18337-0743 MM1 & 4.21 (9) & $41.1^{+1.2}_{-1.1}$ & $9.8^{+0.3}_{-0.3}$ & 485 & 4.2 & $17.1^{+1.0}_{-0.9}$ \\
\multicolumn{6}{c}{} \\
\multicolumn{6}{l}{HMSC} \\
I18151-1208 MM2 & 0.9 (1) & $9.3^{+0.4}_{-0.4}$ & $3.3^{+0.2}_{-0.1}$ & 424 & 2.8 & $17.8^{+1.5}_{-1.4}$ \\
I18182-1433 MM2 & 0.4 (1) & $3.4^{+0.1}_{-0.1}$ & $4.9^{+0.1}_{-0.2}$ & 100 & 0.7 & $13.8^{+1.2}_{-1.1}$ \\
I18223-1243 MM2 & 0.68 (9) & $6.3^{+0.3}_{-0.2}$ & $7.9^{+0.4}_{-0.3}$ & 124 & 0.8 & $15.1^{+1.4}_{-1.3}$ \\
I18223-1243 MM3 & 1.59 (8) & $15.1^{+0.4}_{-0.4}$ & $12.6^{+0.3}_{-0.3}$ & 205 & 1.2 & $16.2^{+1.0}_{-0.9}$ \\
I18223-1243 MM4 & 0.71 (8) & $6.6^{+0.3}_{-0.3}$ & $22.0^{+1.0}_{-1.0}$ & 52 & 0.3 & $15.5^{+1.7}_{-1.5}$ \\
I18306-0835 MM3 & 0.3 (1) & $2.8^{+0.2}_{-0.1}$ & $2.8^{+0.2}_{-0.1}$ & 103 & 1.0 & $16.4^{+3.8}_{-2.9}$ \\
I18337-0743 MM3 & 1.07 (9) & $9.8^{+0.5}_{-0.4}$ & $10.9^{+0.5}_{-0.5}$ & 110 & 0.9 & $15.0^{+1.6}_{-1.4}$ \\
\enddata
\tablenotetext{a}{The errors are written in units of the last significant digit. The numbers in parentheses represent one standard deviation.}
\tablenotetext{b}{The values are taken from \citet{2002ApJ...566..945B} or \citet{2007ApJ...668..348B}.}
\end{deluxetable}


\clearpage

\begin{thebibliography}{}
\bibitem[Alexander et al.(1976)]{1976JMoSp...62...175A} Alexander, A. J., Kroto, H. W., \& Walton, D. R. M.  1976, JMoSp, 62, 175
\bibitem[Bachiller \& P{\'e}rez Guti{\'e}rrez(1997)]{1997ApJ...487L..93B} Bachiller, R., \& P{\'e}rez Guti{\'e}rrez, M.\ 1997, \apjl, 487, L93
\bibitem[Beuther et al.(2002)]{2002ApJ...566..945B} Beuther, H., Schilke, P., Menten, K. M., et al.\ 2002, \apj, 566, 945
\bibitem[Beuther \& Sridharan(2007)]{2007ApJ...668..348B} Beuther, H., \& Sridharan, T. K.\ 2007, \apj, 668, 348
\bibitem[Chapman et al.(2009)]{2009MNRAS.394..221C} Chapman, J. F., Millar, T. J., Wardle, M., Burton, M.~G., \& Walsh, A. J.\ 2009, \mnras, 394, 221
\bibitem[Deleon \& Muenter(1985)]{1985JChPh...82...1702D} Deleon, R. L., \& Muenter, J. S.\ 1985, JChPh, 82, 1702
\bibitem[Dobashi et al.(2005)]{2005PASJ...57S...1D} Dobashi, K., Uehara, H., Kandori, R., et al.\ 2005, \pasj, 57, S1
\bibitem[Foster et al.(2011)]{2011ApJS..197...25F} Foster, J. B., Jackson, J. M., Barnes, P. J., et al.\ 2011, \apjs, 197, 25   
\bibitem[Gerner et al.(2014)]{2014A&A...563A..97G} Gerner, T., Beuther, H., Semenov, D., et al.\ 2014, \aap, 563, A97
\bibitem[Green et al.(2014)]{2014MNRAS.443.2252G} Green, C.-E., Green, J. A., Burton, M. G., et al.\ 2014, \mnras, 443, 2252
\bibitem[Hassel et al.(2011)]{2011ApJ...743..182H} Hassel, G. E., Harada, N., \& Herbst, E.\ 2011, \apj, 743, 182
\bibitem[Herbst \& van Dishoeck(2009)]{2009ARA&A..47..427H} Herbst, E., \& van Dishoeck, E.~F.\ 2009, \araa, 47, 427 
\bibitem[Hirota et al.(2009)]{2009ApJ...699..585H} Hirota, T., Ohishi, M., \& Yamamoto, S.\ 2009, \apj, 699, 585
\bibitem[Hoq et al.(2013)]{2013ApJ...777..157H} Hoq, S., Jackson, J.~M., Foster, J.~B., et al.\ 2013, \apj, 777, 157
\bibitem[Jackson et al.(2013)]{2013PASA...30...57J} Jackson, J. M., Rathborne, J. M., Foster, J. B., et al.\ 2013, \pasa, 30, e057   
\bibitem[Jim{\'e}nez-Serra et al.(2012)]{2012ApJ...753...34J} Jim{\'e}nez-Serra, I., Zhang, Q., Viti, S., Mart{\'{\i}}n-Pintado, J., \& de Wit, W.-J.\ 2012, \apj, 753, 34
\bibitem[Kamazaki et al.(2012)]{2012PASJ...64...29K} Kamazaki, T., Okumura, S. K., Chikada, Y., et al.\ 2012, \pasj, 64, 29 
\bibitem[M$\ddot{\rm u}$ller et al.(2005)]{2005JMoSt...742...215M} M$\ddot{\rm u}$ller, H. S. P., Schl$\ddot{\rm o}$der, F., Stutzki, J., \& Winnewisser, G.\  2005, JMoSt, 742, 215
\bibitem[Nakamura et al.(2015)]{2015PASJ...67..117N} Nakamura, F., Ogawa, H., Yonekura, Y., et al.\ 2015, \pasj, 67, 117 
\bibitem[Nomura \& Millar(2004)]{2004A&A...414..409N} Nomura, H., \& Millar, T. J.\ 2004, \aap, 414, 409
\bibitem[Ohashi et al.(2014)]{2014PASJ...66..119O} Ohashi, S., Tatematsu, K., Choi, M., et al.\ 2014, \pasj, 66, 119
\bibitem[Ohashi et al.(2016)]{2016PASJ...68....3O} Ohashi, S., Tatematsu, K., Fujii, K., et al.\ 2016, \pasj, 68, 3
\bibitem[Sakai et al.(2008)]{2008ApJ...678...1049S} Sakai, T., Sakai, N., Kamegai, K., et al.\ 2008, \apj, 678, 1049
\bibitem[Sakai \& Yamamoto(2013)]{2013ChRv...113...8981} Sakai, N., \& Yamamoto, S.\  2013, ChRv, 113, 8981
\bibitem[Shimajiri et al.(2015)]{2015ApJS..221...31S} Shimajiri, Y., Sakai, T., Kitamura, Y., et al.\ 2015, \apjs, 221, 31
\bibitem[Sridharan et al.(2002)]{2002ApJ...566..931S} Sridharan, T. K., Beuther, H., Schilke, P., Menten, K. M., \& Wyrowski, F.\ 2002, \apj, 566, 931
\bibitem[Sridharan et al.(2005)]{2005ApJ...634L..57S} Sridharan, T. K., Beuther, H., Saito, M., Wyrowski, F., \& Schilke, P.\ 2005, \apjl, 634, L57
\bibitem[Suzuki et al.(1992)]{1992ApJ...392..551S} Suzuki, H., Yamamoto, S., Ohishi, M., et al.\ 1992, \apj, 392, 551
\bibitem[Taniguchi et al.(2016a)]{2016ApJ...817..147T} Taniguchi, K., Ozeki, H., Saito, M., et al.\ 2016a, \apj, 817, 147
\bibitem[Taniguchi et al.(2017a)]{2017ApJ...846...46T} Taniguchi, K., Ozeki, H., \& Saito, M.\ 2017a, \apj, 846, 46
\bibitem[Taniguchi \& Saito(2017)]{2017PASJ...69L...7T} Taniguchi, K., \& Saito, M.\ 2017, \pasj, 69, L7
\bibitem[Taniguchi et al.(2016b)]{2016ApJ...830..106T} Taniguchi, K., Saito, M., \& Ozeki, H.\ 2016b, \apj, 830, 106
\bibitem[Taniguchi et al.(2017b)]{2017ApJ...844...68T} Taniguchi, K., Saito, M., Hirota, T., et al.\ 2017b, \apj, 844, 68
\bibitem[Tatematsu et al.(2010)]{2010PASJ...62.1473T} Tatematsu, K., Hirota, T., Kandori, R., \& Umemoto, T.\ 2010, \pasj, 62, 1473
\bibitem[Tatematsu et al.(2014)]{2014PASJ...66...16T} Tatematsu, K., Ohashi, S., Umemoto, T., et al.\ 2014, \pasj, 66, 16
\bibitem[van der Tak et al.(2007)]{2007A&A...468..627V} van der Tak, F.~F.~S., Black, J.~H., Sch{\"o}ier, F.~L., Jansen, D.~J., \& van Dishoeck, E.~F.\ 2007, \aap, 468, 627
\bibitem[Yamaki et al.(2012)]{2012PASJ...64..118Y} Yamaki, H., Kameno, S., Beppu, H., Mizuno, I., \& Imai, H.\ 2012, \pasj, 64, 118
\bibitem[Yamamoto et al.(1983)]{1983A&A...122...171Y} Yamamoto, T., Nakagawa, N., \& Fukui, Y.\ 1983, \aap, 122, 171
\bibitem[Yu \& Wang(2015)]{2015MNRAS...451...2507Y} Yu, N., \& Wang, J. J.\ 2015, \mnras, 451, 2507
\end{thebibliography}
\end{document}